\documentclass{elsart}

\usepackage{graphicx}

\usepackage{amssymb}

\begin{document}

\begin{frontmatter}



\title{Quarkonium Suppression from SPS to RHIC (and from p+A to A+A)}


\author{Rapha\"el Granier de Cassagnac}
\ead{raphael@in2p3.fr}
\address{Laboratoire Leprince-Ringuet, \'Ecole polytechnique/IN2P3 \\ Palaiseau, 91128 France}

\begin{abstract} Heavy quarkonium production is expected to be
sensitive to the formation of a quark gluon plasma (QGP). It was
(and still is with ongoing data analyses) extensively studied at the
CERN SPS, at collision energy $\sqrt{s_{NN}}$ of the order of
20~GeV. An anomalous suppression was clearly observed. The PHENIX
experiment at RHIC has presented preliminary results that exhibit a
similar amount of $J/\psi$ suppression, at ten times higher
collision energy. I review the results obtained at both facilities.
While interpreting and comparing them, the importance of
understanding normal nuclear effects is emphasized. A new method to
derive a reference for Au+Au collisions from the centrality
dependence of d+Au measurements at RHIC is presented. \end{abstract}


\end{frontmatter}

\section{Nuclear effects from p+A or d+A collisions} \label{Sec:pA}
When trying to interpret all the quarkonia yields observed in p+A
collisions at various energies and kinematical domains, we find
ourselves facing a real puzzle~\cite{Mike:HP04}. Various effects are
invoked, including normal nuclear absorption of quarkonia or
pre-resonant $c\overline{c}$ pairs, parton shadowing and
corresponding anti-shadowing, contribution from the intrinsic charm
existing in the nucleon wave function, or energy loss and related
transverse momentum broadening, all of them being further
complicated by feed-down from higher mass states. Explaining
all the available p+A data is beyond the scope of this article. In
the following, I restrict my interest for p+A data and nuclear
effects to the energies and kinematical domains where A+A collisions
are also measured, in order to set up references for QGP studies.

\subsection{Normal nuclear absorption at SPS}

\begin{figure}
\begin{center}
  \begin{tabular}{cc}
  \includegraphics[width=7cm]{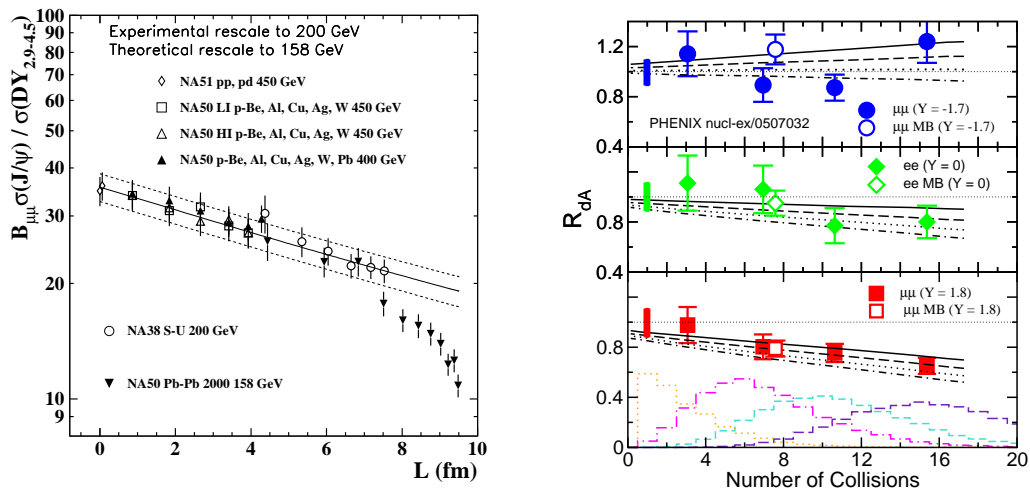} & \includegraphics[width=6.2cm]{Fig_RdA_ncoll.eps}
  \end{tabular}
\end{center} \caption{Normal nuclear matter effects. Left) At SPS, the
$J/\psi$/Drell-Yan cross-section ratio versus the average nuclear
length $L$, for several collision systems~\cite{NA50:final}. The
line stands for a normal nuclear absorption of $\sim 4.2$~mb. Right)
At RHIC, $J/\psi$ nuclear modification factor $R_{dA}$ as a function
of the number of nucleon-nucleon collisions $N_{coll}$ for backward
(top, $y=-1.7$) mid (middle, $y=0$) and forward (bottom, $y=1.8$)
rapidities~\cite{phenix:dA}.
Theoretical curves from Vogt~\cite{Vogt:dA}, assuming EKS
shadowing and 0, 1, 2 or 3~mb normal absorption cross-sections (from
top solid to bottom dot-dashed).} \label{Fig1} \end{figure}

Figure \ref{Fig1} (left) summarizes the published $J/\psi$
production yield (normalized to Drell-Yan) observed at SPS energies,
from p+p to Pb+Pb collisions~\cite{NA50:final}, excluding In+In data
from the NA60 experiment~\cite{NA60}. It is plotted here as a
function of the average length $L$ of nuclear matter traversed by
the $c\overline{c}$ state. Looking only at p+A collisions, we
observe a clear exponential suppression. This behavior is expected
if the only nuclear effect is an absorption by the nucleons in the
incoming nuclei. The line of figure~\ref{Fig1} (left) stands for
such a normal nuclear absorption, with a fitted $J/\psi$-nucleon
cross-section of $\sigma^{J/\psi}_{abs}=4.18 \pm 0.35$~mb. Although
this simple picture does a splendid job in reproducing the data, all
the way from p+p up to peripheral Pb+Pb collisions (including the
whole S+U range), one can oppose theoretical arguments against it.
First, the p+A data are rescaled to a lower collision energy
$\sqrt{s_{NN}}$ (from 27 or 29~GeV to 17 and 19~GeV). The NA60
collaboration is currently analyzing p+A data at 17~GeV. Second,
according to various theoretical predictions~\cite{shadow},
shadowing (or rather anti-shadowing) could play a role at the SPS
experiments regime, namely for momentum fraction $x$ of the order of
$10^{-1}$. The SPS experiments can hardly address this question
because of their limited rapidity (or $x$) range. Still, the
rapidity asymmetry observed in $J/\psi$ yields in p+A collisions
might be a hint of such an effect~\cite{NA50:pA}. Third, the p+A
absorption might not be straightly applicable to A+A, for instance
if there are complications due to changing feed-down ratios. Indeed,
the $\psi'$ is known not to suffer the same nuclear effects as the
$J/\psi$: its absorption cross-section extracted from p+A is
$\sigma^{\psi'}_{abs}=7.7 \pm 0.9$~mb and its anomalous suppression
already sets in S+U collisions~\cite{NA50:pA}. However, one should
not forget that the simple picture of nuclear absorption perfectly
describes a large amount of data (p+p, various p+A, S+U and
peripheral Pb+Pb). The departure from the absorption curve for more
central Pb+Pb collisions is clearly a new phenomenon.

\subsection{Shadowing and absorption at RHIC}

At $\sqrt{s_{NN}}=200$~GeV, the only available data on quarkonia
from \emph{p+A like} collisions are $J/\psi$ seen by the PHENIX
experiment in d+Au~\cite{phenix:dA}, probing a wide rapidity range,
from $-2.2$ to 2.4, corresponding to momentum fractions $x$ of
gluons in the gold nuclei ranging from $\sim 10^{-3}$ to $\sim
10^{-1}$. The minimum bias points of figure~\ref{Fig1} (right) show
that the measured nuclear modification factors $R_{dA}$ depend on
rapidity. This is interpreted as due to shadowing and/or
anti-shadowing. While the strength of gluon shadowing is not heavily
constrained by theory (models predictions \cite{shadow} differ by a
factor of three), PHENIX data favor moderate shadowing schemes such
as the Eskola-Kolhinen-Salgado (EKS). In addition to this, a
moderate normal nuclear absorption is allowed, not larger than 3~mb.
The addition of these two ingredients is performed by Vogt
in~\cite{Vogt:dA} and can describe both the rapidity and centrality
dependencies, as shown on figure~\ref{Fig1} (right).


\subsection{From d+Au to Au+Au at RHIC}

\begin{figure}
\begin{center}
  \begin{tabular}{cc}
  \includegraphics[width=6.5cm]{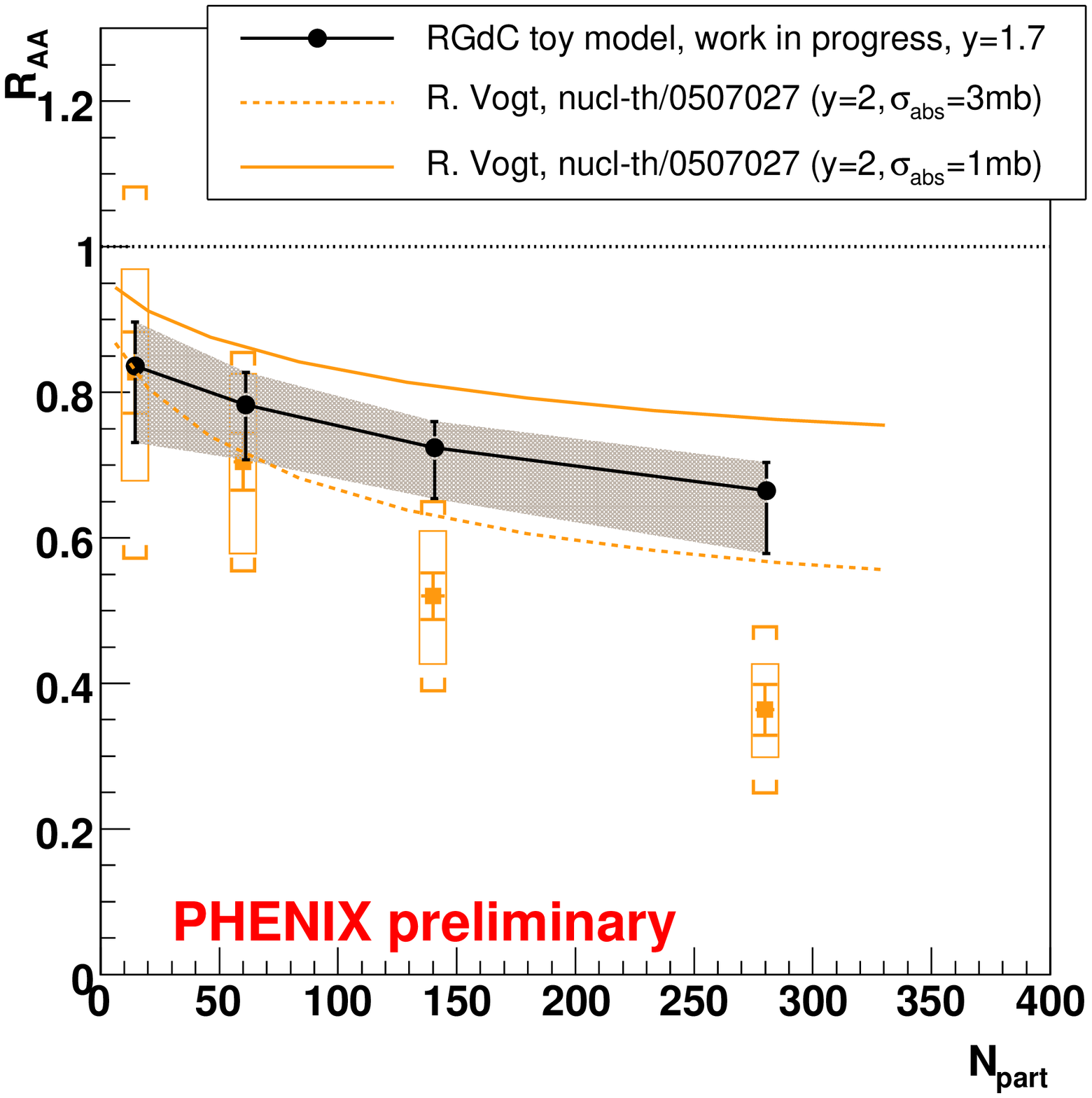} & \includegraphics[width=6.5cm]{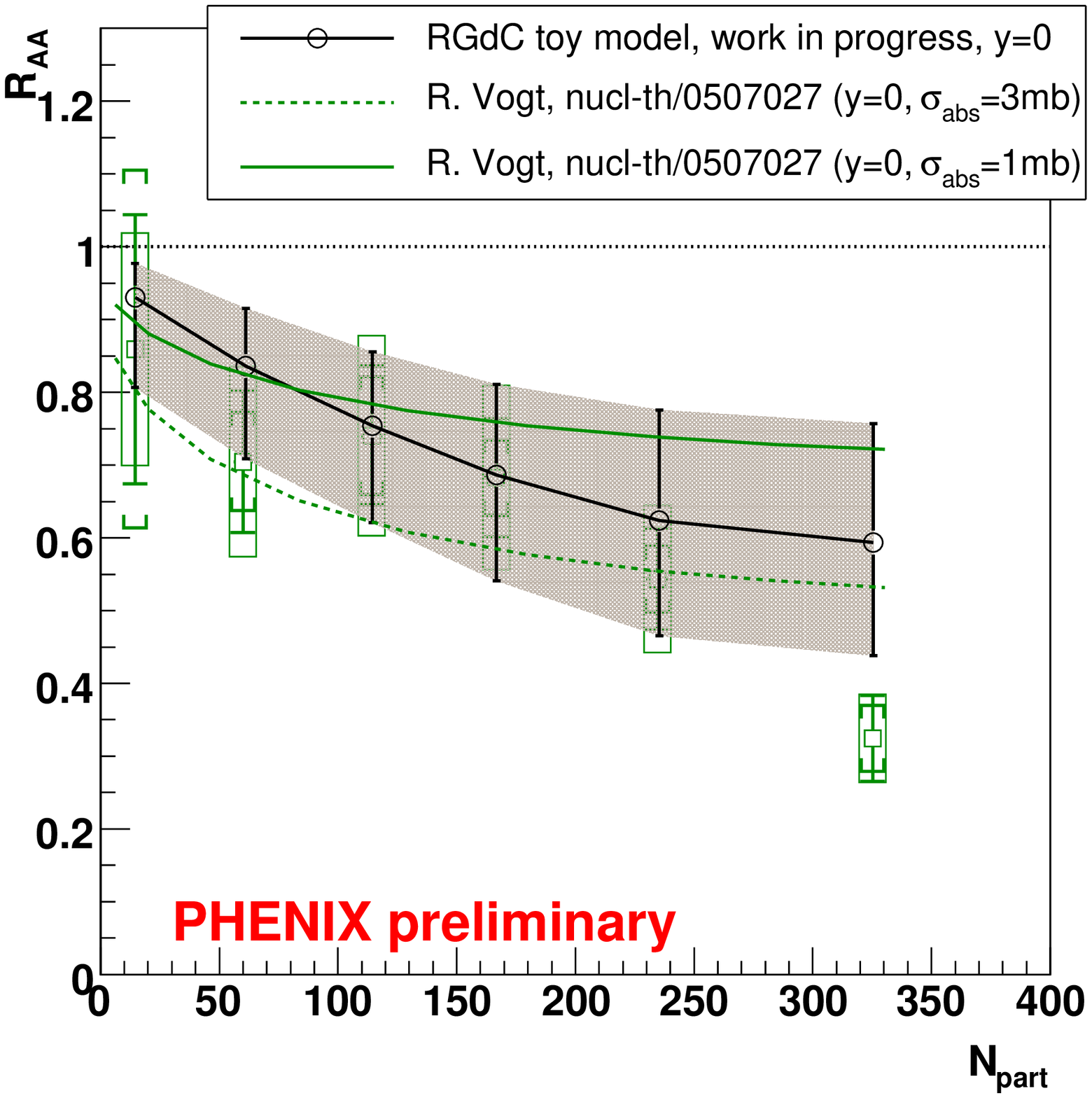}
  \end{tabular}
\end{center} \caption{Nuclear modification factor $R_{AA}$ as a
function of centrality (given here by the number of participants
$N_{part}$) for Au+Au at $y=1.7$ (left) and $y=0$ (right). Squares
are preliminary data from PHENIX~\cite{phenix:AA}.
Theoretical curves are nuclear effects predictions from
Vogt~\cite{Vogt:AA}, solid and dashed lines being for 1 and 3~mb
normal nuclear absorption cross sections, respectively. The circles
within the shaded bands show the prediction from the model presented
here.} \label{Fig2}
\end{figure}

To interpret $J/\psi$ production in A+A collisions at RHIC, we need
a model capable of reproducing the d+Au nuclear modification factors
$R_{dA}$ and extrapolating them to A+A. Such a model, including
inhomogeneous shadowing and nuclear absorption is given
by~\cite{Vogt:AA} and shown on figure~\ref{Fig1} (right).
Another attempt to derive a reference from d+Au can be found
in~\cite{KKS} where the authors derive dissociation cross-sections
from the centrality dependence of $R_{dA}$. They assume that nuclear
effects are proportional to $\exp(-\rho_0 \sigma_{diss} L)$,
$\rho_0$ being the normal nuclear density and $L$ the average length
of nuclear matter seen by the $J/\psi$. However, there is no
fundamental reason for this function to reflect the centrality
dependence of shadowing. I propose here an alternate method, with a
concern to be as much data-driven as possible. First, I perform
phenomenological fits of the modification factors $R_{dA}(y,b)$ as a
function of the impact parameter $b$ (given by a Glauber model), for
the three rapidities $y$ of the PHENIX measurements. Given the
experimental uncertainties, linear fits are sufficient to describe
the data. I then run a A+A Glauber model. For each A+A collision
occurring at a given impact parameter $b_{AA}$, the positions of the
$N_{coll}$ elementary nucleon+nucleon collisions are randomly
distributed (following the nuclear densities) providing the
locations $b^i_1$ and $b^i_2$ of each collision $i$, relative to the
center of nucleus 1 and nucleus 2. For the considered A+A collision,
the predicted nuclear modification factor is given by the following
summation over the elementary collisions: $ R_{AA} (|y|, b_{AA}) =
\sum^{N_{coll}}_{i=1} ( R_{dA}(-y,b^i_1) \times R_{dA}(+y,b^i_2) ) /
N_{coll}$. This formula assumes that a $J/\psi$ produced in a A+A
collision at a given rapidity $y$ suffers the product of the nuclear
effects that were observed in d+A at this rapidity, by the ones of
the opposite rapidity\footnote{A similar assumption is made in
\cite{KKS} while summing dissociation cross-sections.}(equivalent to
a A+d collision). This assumption is correct for the only two
effects considered so far to explain RHIC data, namely shadowing and
nuclear absorption. Quarkonia production is proportional to the
parton distribution functions ($pdf$) in each nucleus, while the
average length is the sum of the length in each nucleus, so that the
production is finally proportional to $pdf_1 \times pdf_2
\times \exp (-\rho \sigma (L_1+L_2))$. This method has two
advantages\footnote{Despite these advantages it will be difficult to
apply it at LHC where p+A and A+A collisions are planed to be
measured at difference collision energies and $J/\psi$ in ALICE
 will only be measured at positive \emph{or} negative rapidity for p+A.}. First, the
statistical and systematical uncertainties of the d+A measurements
can be directly propagated to the A+A prediction. Second, since it
is based on a Glauber calculation, it is easy to predict
modification factors for experimental centrality classes. This is
done on figure~\ref{Fig2} where predictions are given for the Au+Au
PHENIX centrality classes. No systematic uncertainties from the
method itself (Glauber parameters) are calculated. The amount of
predicted suppression is compatible with Vogt's predictions. In the
forward rapidity case (left) its uncertainty is smaller than the
allowed variation between the 1 and 3~mb absorption cross-sections
and seem to favor intermediate ones. In the mid-rapidity case (right) the
uncertainty is larger. At both rapidities, the $J/\psi$ suppression
seen by PHENIX~\cite{phenix:AA} in the most central data is larger
than the one predicted by Vogt or by the model described above,
pointing out that there is an anomalous suppression at RHIC energy.


\section{Anomalous suppressions} \label{Sec:AA}

In both SPS (figure \ref{Fig1} left) and RHIC (figure \ref{Fig2})
data, the most central A+A $J/\psi$ measurements depart from the
nuclear effect predictions, suggesting that other mechanisms are
involved. Such an \emph{anomalous} suppression was early predicted
by Matsui and Satz as a signature of the QGP~\cite{Satz}.

\subsection{From SPS to RHIC}

\begin{figure}
  \begin{tabular}{cc}
  \includegraphics[width=6.5cm]{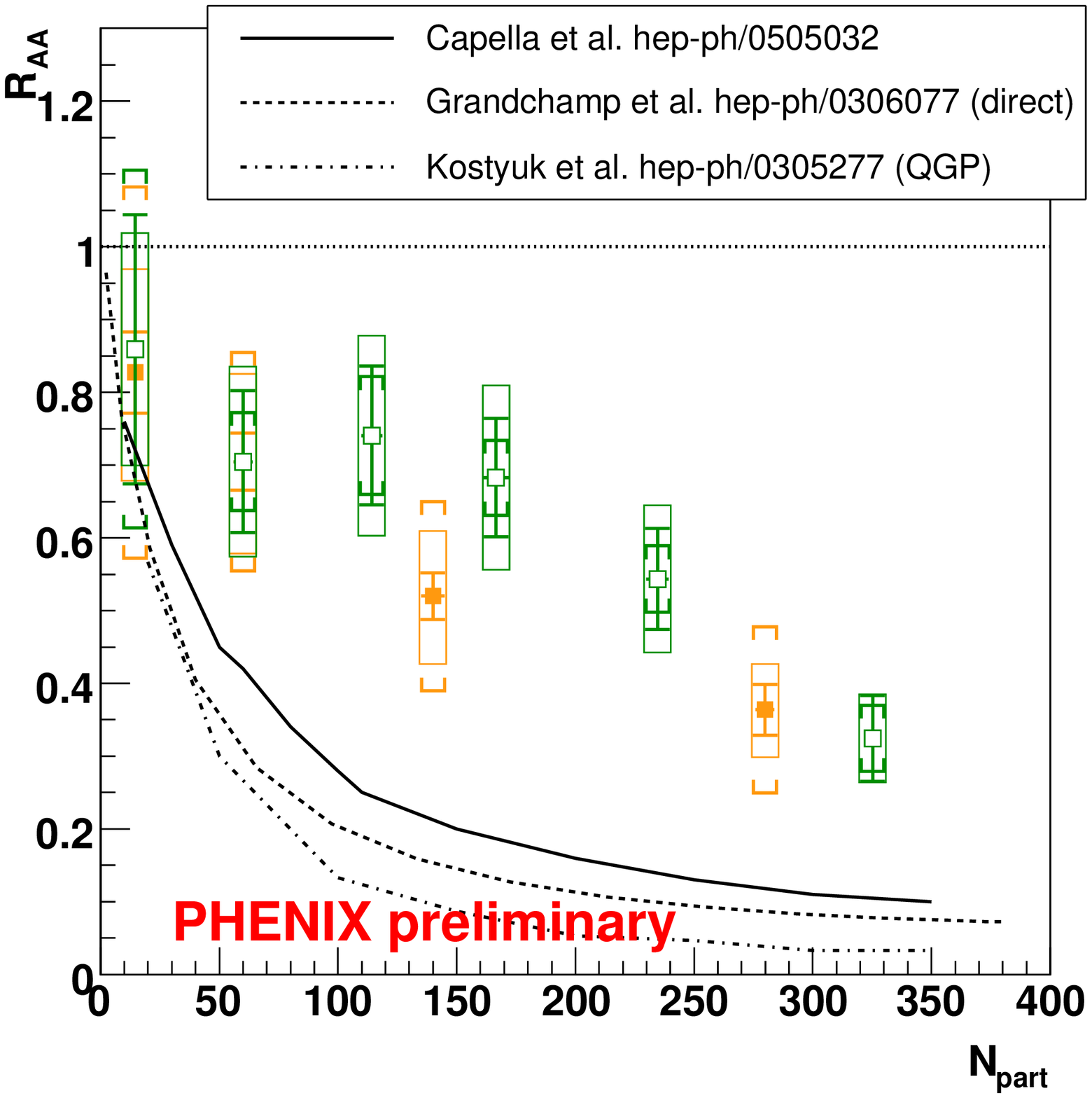} & \includegraphics[width=6.5cm]{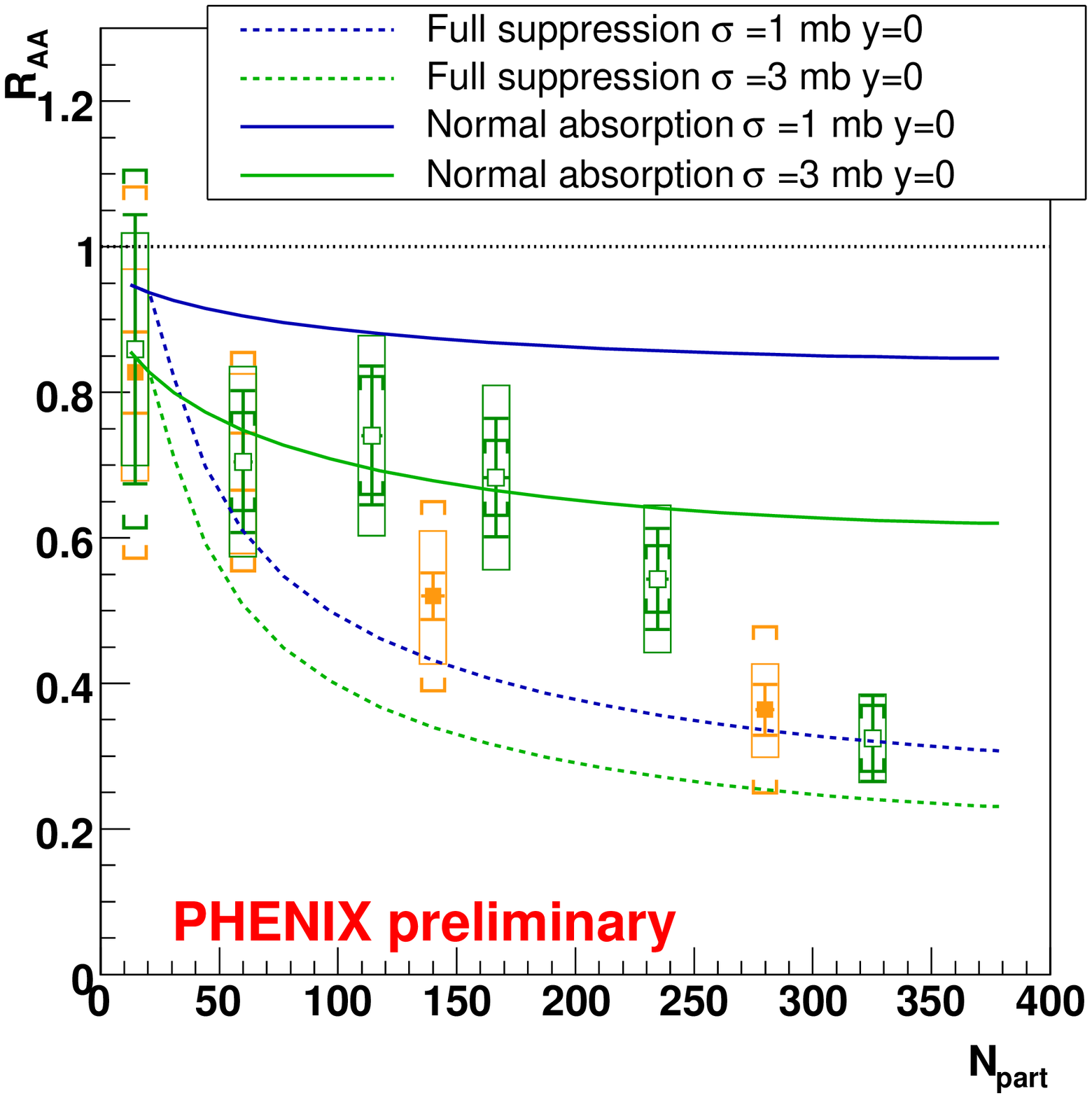}
  \end{tabular}
\caption{RHIC Nuclear modification factor $R_{AA}$ as already presented on
figure~\ref{Fig2} versus predictions derived from models reproducing
SPS data. Theoretical curves are predictions from models that
describe the SPS anomalous suppression. Left) Most of the models
over-predict the suppression~\cite{Capella,Grandchamp,Kotsyuk}.
Right) One model~\cite{Zhu} reproduces the most central suppression
(solid/dashed are without/with plasma, top/bottom assume 1/3~mb
normal absorption cross-section).} \label{Fig3}
\end{figure}

Various models account for the $J/\psi$ anomalous suppression seen
in Pb+Pb collisions at SPS. They were used to predict the expected
$J/\psi$ yield at RHIC energy. Three of these predictions are
compared to PHENIX Au+Au preliminary data~\cite{phenix:AA} on
figure~\ref{Fig3} (left). In
\cite{Capella} (solid line), $J/\psi$'s are absorbed by comoving
particles (of undetermined partonic/hadronic nature). In
\cite{Grandchamp}, the authors describe the dynamical interplay
between suppression and regeneration of $J/\psi$'s in a QGP. The
suppression mechanism is dominant for NA50 energies and is the only
one presented here as a dashed line (see figure \ref{Fig4} for the
full prediction). In
\cite{Kotsyuk} (dot-dashed line), a QGP statistical charm
coalescence model is used. All three models fail to reproduce PHENIX
data, overestimating the measured suppression. Other models such as
percolation~\cite{Percolation} also over-predicts the suppression,
suggesting that new mechanisms take place at RHIC energy. It is
interesting to note that the same models also failed to reproduce
the In+In data shown by the NA60 experiment~\cite{NA60}.

\subsection{Alternate explanations for RHIC suppression}

Three classes of models exist that can accommodate the amount of
anomalous suppression seen in the most central collisions. The
accuracy of present data and the nuclear effects uncertainty, do not
allow to favor one over the other.

\begin{figure}
  \begin{tabular}{cc}
  \includegraphics[width=6.5cm]{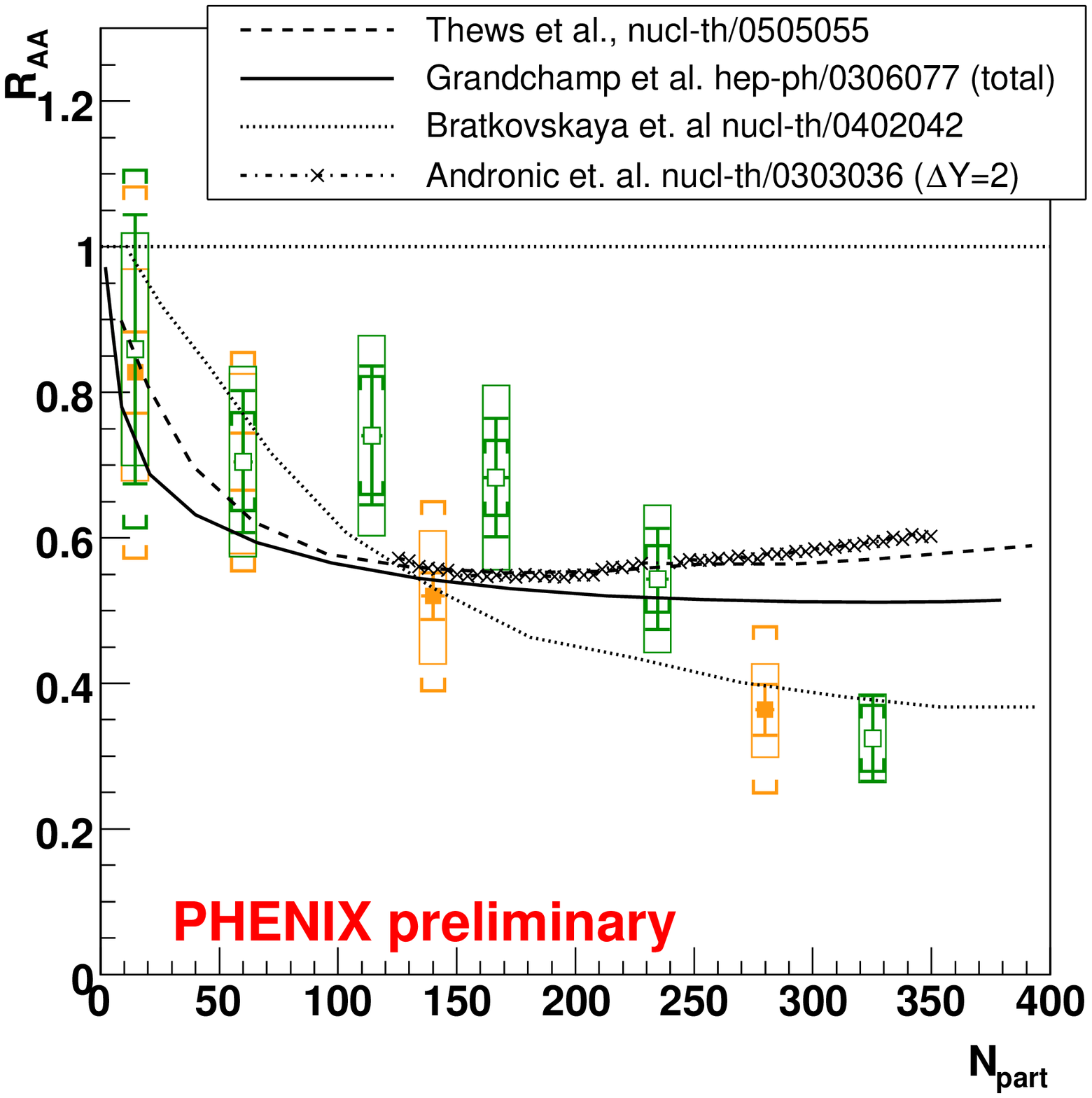} & \includegraphics[width=6.5cm]{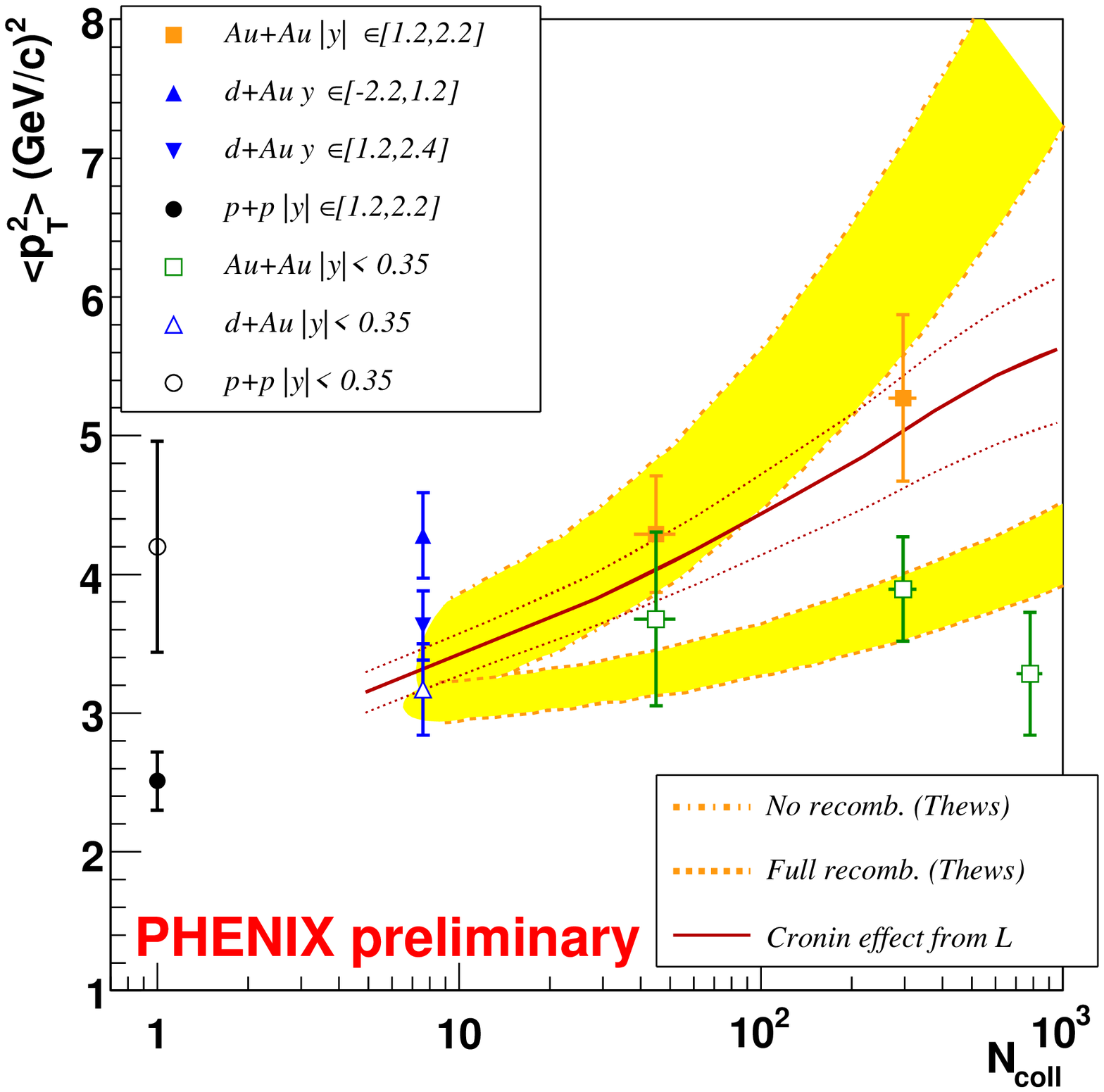}
  \end{tabular}
 \caption{Recombination scenarios at RHIC. Left) Nuclear modification factor $R_{AA}$ as already presented
on figure~\ref{Fig2}. Theoretical curves come from various
recombination models \cite{Grandchamp,Bratkoskaya,Andronic,Thews}.
Right) Mean squared transverse momentum versus $N_{coll}$, from p+p
(circles) to d+Au (triangles) and Au+Au (squares) for forward/mid
rapidities (open/full symbols). Upper/lower shaded bands
from~\cite{Thews} stand for direct/fully recombined $J/\psi$. The
solid line is a parametrization of Cronin effect derived from d+Au
data in~\cite{Tram}, the dotted lines reflecting the associated
errors. } \label{Fig4}
\end{figure}

 \begin{itemize}
 \item{Detailed transport:}
One paper~\cite{Zhu}, simulating $J/\psi$ transport in a
hydrodynamical model, predicts an amount of suppression that matches
the most central data. It is shown on figure~\ref{Fig3} (right)
where the authors have added nuclear matter effects (nuclear
absorption only, 1 or 3~mb) with respect to the published paper. The
suppression they obtain is not large probably because of their
description of the boundary between the QGP and the nuclear phase.


\item{Sequential melting:} An important fraction (30 to 40\%) of
$J/\psi$'s comes from decays of excited states ($\psi'$, $\chi_c$)
as it is shown by the HERA-B experiment~\cite{HERA-B}. They are
taken into account in most of the approaches. Recent lattice
computations indicate that $J/\psi$'s could melt at a much higher
temperature than the one that was originally thought. One possible
hypothesis, defended in~\cite{KKS}, is that, both at SPS and RHIC,
only the excited states melt, leaving all the initially produced
$J/\psi$'s untouched.

\item{Recombination:} At RHIC energies, multiple $c\overline{c}$
pairs are produced, 10 to 20 in central
collisions~\cite{phenix:charm}. Quark mobility in a deconfined
medium could allow uncorrelated charm quarks to recombine when the
QGP fireball freezes, raising the quarkonium yield with centrality.
A balance between suppression and enhancement could lead to the
intermediate suppression observed at RHIC. Figure~\ref{Fig4} (left)
shows a collection of predictions from various recombination or
coalescence models~\cite{Grandchamp,Bratkoskaya,Andronic,Thews}.
Unfortunately, the lack of knowledge concerning yields and
distributions of
 the initially produced charm quarks, as well as of the recombination
mechanism, make these predictions hardly predictive. A way to search
for recombination is to look at its impact on the distributions of
kinematical variables, such as transverse momentum.
\end{itemize}


\section{$J/\psi$ mean transverse momentum}
\label{sec:MeanPt}

Recombination is expected to modify transverse momentum
distributions. To properly predict the modified $p_T$ spectra, one
first needs to quantify the $p_T$ broadening coming from normal
nuclear effects (Cronin effect). This effect was clearly seen at SPS
by comparing p+p and p+A $\langle p_T^2 \rangle$, as well as in
PHENIX at forward\footnote{The midrapidity p+p has too poor
statistics to claim for a modification with respect to the d+Au
measurement.} rapidity~\cite{phenix:dA}. At SPS, a simple
parametrization could reproduce the $\langle p_T^2 \rangle$ values
from p+p up to Pb+Pb: $
\langle p_T^2 \rangle_{AA} = \langle p_T^2 \rangle_{pp} + \rho
\sigma \delta(\langle p_T^2 \rangle) \times L $ where $L$ is the
average thickness of nuclear matter seen by a $J/\psi$. The factor
$\rho \sigma \delta(\langle p_T^2 \rangle)$ stands for the nuclear
density $\rho$, times the elastic gluon-nucleon scattering cross
section $\sigma$, times the average $p_T$ kick given at each
scattering $\delta(\langle p_T^2 \rangle)$. A review of SPS
(including some FNAL data) was made in~\cite{Tram} together with a
similar fit to RHIC forward data. This is presented as a solid line
(with associated errors) on figure~\ref{Fig4} (right). The shaded
bands are predictions from~\cite{Thews}, corresponding to either
$J/\psi$'s from recombination (lower band) or to directly produced
$J/\psi$'s (upper band). No clear sign for modification is seen and
the A+A $\delta(\langle p_T^2 \rangle)$ can be interpreted in terms
of normal broadening.

The rapidity spectra are also expected to be modified by
recombination, but there is also no sign of this so
far~\cite{phenix:AA}.



As a conclusion, I stress that RHIC and SPS data are not so easy to
compare, even if they exhibit similar suppression at their highest
energy densities. The amount of normal nuclear suppression is poorly
known, especially at RHIC where it demands more d+A data.
Nevertheless, RHIC preliminary suppression seems anomalous and all
the different models that can accommodate it suppose the formation
of a QGP. To distinguish between them, a better precision on data,
and in particular on the kinematical distributions, is required.


\begin{thebibliography}{00}

\bibitem{Mike:HP04} M.J. Leitch, Eur. Phys. J. \textbf{C43} (2005),
157-160.

\bibitem{NA50:final} NA50 collaboration, Eur. Phys. J. \textbf{C39}
(2005), 335-345.

\bibitem{NA60} NA60 collaboration, R. Arnaldi, this conference.

\bibitem{shadow} K.~Eskola, V.~Kolhinen, and C.~Salgado, Eur. Phys.
J. \textbf{C9} (1999) 61-68, L.~Frankfurt and M.~Strikman, Eur.
Phys. J. \textbf{A5} (1999) 293-306, B.~Kopeliovich, A.~Tarasov et
J.~H\"ufner, Nucl. Phys. \textbf{A696} (2001) 669-714.



\bibitem{NA50:pA} NA50 Collaboration, CERN-PH-EP/2006-018, Eur. Phys. J. C., in print.

\bibitem{phenix:dA} PHENIX collaboration, Phys. Rev. Lett.
\textbf{96}, (2006) 012304.

\bibitem{Vogt:dA} R. Klein and R. Vogt, Phys. Rev. Lett.
\textbf{91}, (2003) 142301.


\bibitem{Vogt:AA} R. Vogt, nucl-th/0507027.

\bibitem{KKS} F. Karsch, D. Kharzeev and H. Satz, Phys. Lett.
\textbf{B637} (2006) 75-80.

\bibitem{phenix:AA} PHENIX collaboration, H. Pereira da Costa, Nucl.
Phys. \textbf{A774} (2006) 747-750, A. Bickley, this conference.

\bibitem{Satz} T. Matsui and H. Satz, Phys. Lett. \textbf{B178}
(1986) 416.

\bibitem{Capella} A. Capella and E.G. Ferreiro, Eur. Phys. J.
\textbf{C42} (2005) 419-424.

\bibitem{Grandchamp} L. Grandchamp et al., Phys. Rev. Lett.
\textbf{92} (2004) 212301.

\bibitem{Kotsyuk} Kostyuk et al., Phys. Rev. \textbf{C68} (2003)
041902. 

\bibitem{Percolation} S. Digal et al., Eur. Phys. J. \textbf{C32}
(2004) 547-553.

\bibitem{Zhu} X.L. Zhu et al., Phys. Lett. \textbf{B607} (2005)
107-114.

\bibitem{HERA-B} HERA-B collaboration, P. Faccioli, this conference.

\bibitem{phenix:charm} PHENIX collaboration, Phys. Rev. Lett.
\textbf{94} (2005) 082301.

\bibitem{Bratkoskaya} E.L. Bratkovskaya et al., Phys. Rev.
\textbf{C69} (2004) 054903.

\bibitem{Andronic} A. Andronic et al., Phys. Lett. \textbf{B571}
(2003) 36-44.

\bibitem{Thews} R.L. Thews and M.L. Mangano, Phys. Rev. \textbf{C73}
(2006) 014904.

\bibitem{Tram} V.N. Tram, PhD thesis, \'Ecole polytechnique and
nucl-ex/0606017.




\end{thebibliography}
\end{document}